%% file: main.tex
\documentclass{article}
\usepackage[
    backend=biber,
    style=nature,
  ]{biblatex}

\usepackage[utf8]{inputenc}
\usepackage[T1]{fontenc}
\usepackage{url}
\usepackage{fullpage}
\usepackage{amsthm,amsmath,amssymb,microtype,graphicx,subcaption,booktabs,enumitem,hyperref,rotating,caption,calc}
\usepackage[capitalise]{cleveref}
\usepackage{tikz}
\usepackage[normalem]{ulem}
\usepackage{authblk}

\DeclareCaptionStyle{figstyle}
  [format=plain,margin=0pt,justification=centering]
  {format=plain,margin=8pt}
\input{tex/preamble}

\begin{document}

\title{Verifiable Proof of Health using Public Key Cryptography}

\author[1]{Abhishek Singh}
\author[1]{Ramesh Raskar}
\affil[1]{\footnotesize MIT Media Lab}

\maketitle

\begin{abstract}
In the current pandemic, testing continues to be the most important tool for monitoring and curbing the disease spread and early identification of the disease to perform health-related interventions like quarantine, contact tracing and etc. Therefore, the ability to verify the testing status is pertinent as public places prepare to safely open. Recent advances in cryptographic tools have made it possible to build a secure and resilient digital-id system. In this work, we propose to build an end to end COVID-19 results verification protocol that takes privacy, computation, and other practical concerns into account for designing an inter-operable layer of testing results verification system that could potentially enable less stringent and more selective lockdowns. We also discuss various concern encompassing the security, privacy, ethics and equity aspect of the proposed system.
\end{abstract}

\input{tex/intro}

\input{tex/rw}

\input{tex/method}

\input{tex/analysis}

\input{tex/discussion}

\input{tex/conclusion}


\printbibliography
\end{document}

%% file: tex/preamble.tex
\newcommand{\KeyGen}{\mathsf{KeyGen}}
\newcommand{\Encrypt}{\mathsf{Encrypt}}

\newcommand{\Decrypt}{\mathsf{Decrypt}}
\newcommand{\pk}{\mathsf{pk}}
\newcommand{\sk}{\mathsf{sk}}
\newcommand{\Apk}{\mathsf{A_{pk}}}
\newcommand{\Ask}{\mathsf{A_{sk}}}
\newcommand{\phask}{\mathsf{PHA_{sk}}}
\newcommand{\phapk}{\mathsf{PHA_{pk}}}

%% file: tex/intro.tex
\section{Introduction}\label{sec:intro}
The economy is one of the sector that has suffered enormous negative impact during this pandemic. One of the major reason for such a disruption is the lockdown policy enforced by governments causing all the public places to shut down. While such a policy is important for curbing the disease spread, its side effects manifest in different forms including financial loss, mental health, economic collapse and etc. In this work, we describe a cryptographic scheme through which citizens, healthcare authorities, and public place owners can work collaboratively by restricting access to infectious individuals for public places. Such a scheme would allow less stringent and more selective lockdowns and a potential reduction in the disease spread by cutting down infection chain at the places of public gathering. Previous works~\cite{chang2020mobility,qian2020indoor} in the analysis of disease spread have shown the possibility of such public places turning into a super spreading zone. Therefore, such a health passport would benefit the current overarching goal of managing the economy and disease spread simultaneously. That being said, it is important to highlight that the design and deployment of such a technique should be done after discussion with various experts including policy makers, epidemiologists, privacy experts, and ethicists. Hence, the scope of this paper is only about the design of the cryptographic scheme and not a final end to end solution.\\
There has been a reasonable number of prior work in this area that aim to build a similar health passport for verified access. A majority of them borrow technology from the blockchain technology and verifiable credentials. However, as we show in our section~\ref{sec:method}, our proposed protocol does not require any heavyweight computing infrastructure and it requires a minimal communication with the internet for the protocol to operate successfully. One of the drawback of such a design is its limited capability to only verify a testing or immunization result and not support any arbitrary credential verification or transaction, which we consider is sufficient under the scope of the current pandemic.\\
The proposed protocol in this work offers threefold advantage:
\begin{itemize}
    \item It uses simple widely known yet secure cryptographic primitives.
    \item It is designed to be integrated with the existing testing ecosystem in an inter-operable fashion.
    \item The protocol addresses the challenges that are relevant to the practical deployment such as offline verification and health status forgery.
\end{itemize} 

%% file: tex/rw.tex
\section{Related Work}\label{sec:rw}
There has been a significant work in the domain of verifiable ecosystem of credentials using digital tools. The most notable work is the verifiable credential standard described by W3C~\cite{W3C}. Many existing frameworks utilize the similar setup of verifiable credential~\cite{angelopoulos_dhp_2005}. COVID-19 credentials initiative~\cite{noauthor_covid-19_nodate} is a consortium of multiple big companies and groups working in the domain of verifiable credential working towards building technology around addressing some of the use-cases for COVID-19. There are many other frameworks proposed recently~\cite{noauthor_validatedid_nodate, butler_differentially_2020,chaudhari_framework_2020,hicks_secureabc_2020,angelopoulos_dhp_2005-1,eisenstadt_covid-19_2020}. However, this line of work also comes with criticism~\cite{noauthor_vision_nodate}. While our work is inspired by all of these initiatives, the key difference is the underlying protocol used in our work that does not require any distributed ledger or a verifiable credential type of setup but simply relies on public key cryptography. Furthermore, we aim to address some of the practical problems associated with these verifiers in a realistic setting such as offline verification, integration with testing site and the PHA server for a given jurisdiction. We are able to obtain such advantages without using any sophisticated cryptographic primitives because of the assumption of tightly couple ecosystem which these verifiable credentials based work do not assume.\\
This paper draws some of the ideas from the work done in cryptography for message authentication and signatures~\cite{vidakovic2013rsa,nyberg1994message,pornin2013deterministic}. There are several protocols that allow designing authentication to message and signature schemes such as HMACs~\cite{krawczyk1997hmac}. However, these works do not suffice for preventing forgery while also performing offline verification. Here, by forgery, we refer to the situation where multiple users carry same verified information and show the same verifiable proof even though it does not belong to them. Either forgery is resolved by exchanging message with the \textit{issuer} or by offline authentication algorithms where message. We solve both of the problem in our protocol by allowing the server to embed user identifier (user's public key) in the signed copy of result. This prevents the forgery of the message as long as the private key of the user is not leaked.

%% file: tex/method.tex
\section{Methodology}\label{sec:method}
\subsection{Problem Definition}
At a high level, a health based verification system involves three parties \textit{issuer}, \textit{holder}, and \textit{verifier}. For the use-case of COVID-19, the \textit{issuer} is a public healthcare authority (PHA) that usually sets up a \textit{testing site} and issues the test results to individuals. The individuals here act as \textit{holder} of the result and want the access to public places based on their health status. Finally, the \textit{verifier} verifies the health status of the \textit{holder} by inspecting the health status of the individuals. The goal of the \textit{verifier} is to only authorize access to those individual who satisfy the following requirements -
\begin{itemize}
    \item Carry a \textit{health status} that belongs to \textit{holder}
    \item The \textit{health status} should be authorized by \textit{issuer}
    \item The \textit{health status} contains appropriate result(ex - negative test in the last 7 days).
\end{itemize}
\subsection{Terminologies}
We use $\oplus$ to denote the XOR operation. Whenever a tuple of data is exchanged, all of the elements of the tuple are in the format $(x,y,z)$ where $x,y$ and $z$ are separate messages to be exchanged. Under the given setup we refer the \textit{holder} as $A$ and \textit{issuer} as \textit{PHA}.
\subsection{Requirements}\label{sec:req}
\paragraph{Functional}
\begin{itemize}
    \item F1: The \textit{verifier} should be able to verify the \textit{health status} without requiring an internet connection.
    \item F2: The \textit{holder} should be able to present the \textit{health status} without requiring an internet connection.
    \item F3: The verification process should be quick and computationally efficient.
    \item F4: No two users should be able to share the \textit{health status} with each other and gain the access through verifier fraudulently.
\end{itemize}
\paragraph{Privacy}
\begin{itemize}
    \item P1: The \textit{verifier} should only be able to query the \textit{health status} of an individual that presents \textit{health status} with consent.
    \item P2: The \textit{verifier} should not be able to learn any other information about the \textit{holder} except their health status.
    \item P3: Any individual should not be able to query the \textit{health status} of any arbitrary user except only the user who queries its own entry.
    \item P4: Existence of an individual on the public health database should not be identifiable.
\end{itemize}
\subsection{Threat Model}
There are two categories of threat at a high level in this proposed protocol. The first category refers to forgery where a user attempts to bypass the verification. This could happen by either altering the content in \textit{result status} $R$ (modifying positive test as a negative test or changing the last tested timestamp to a permissible one). This can also happen if user can obtain someone else's $R$. The second category is associated with privacy. Under this threat model the \textit{verifier} acts as a semi-honest adversary attempting to reveal more information about the \textit{holder}. This could either happen through revealing more information during \textit{verification} stage or post \textit{verification stage} where the \textit{verifier} attempts to query the data on server through information learned during \textit{verification}. Under our threat model, no two parties are colluding with each other at any point of time. However, any arbitrary number of \textit{verifiers} or \textit{holders} can collude among each other to leak additional information \textit{holder} or to bypass \textit{verifier} respectively. Furthermore, in this proposed protocol, the \textit{issuer} is a trusted entity which is an obvious assumption to make because the \textit{issuer} knows the result and identity of every \textit{holder} from the beginning of the testing phase and for the scope of this paper, \textit{issuer} is Public Healthcare Authority. In line with that, the protocol assumes that \textit{issuer's} public key $\phapk$ is known to everyone.

\subsection{Preliminaries}
While the proposed protocol is agnostic to the underlying encryption algorithm as long as it supports public and private key cryptography. For the sake of description, we base our system on el-gamal~\cite{elgamal} which is a well known public key crypto system. We define the protocol as follows:
\begin{itemize}
    \item $\KeyGen(G, q, g) \rightarrow \sk, \pk$: Construct group $G$ of order $q$ with generator $g$. The private key is an integer $e$ randomly sampled from the group $G$. The public key is $g^e$ and the group $G$.
    \item $\Encrypt(m, \sk) \rightarrow (c_1, c_2)$: a third party can encrypt a message $m$ by first sampling random $d$ from the group $G$ and computing $c_1 \equiv g^d$ and $c_2 \equiv m\cdot(g^e)^d$.
    \item $\Decrypt((c_1, c_2), \pk) \rightarrow m$: is performed by the receiving party only if they know the private key $e$ as $m\equiv (c_2)\cdot((c_1)^e)^{-1} (mod(n))$. The inverse can be easily computed by using the value of $q$.
\end{itemize}
 While the above definition suits well for the protocol, there are known issues with the plain El-gamal system~\cite{elgamal} for which it could be substituted with secure and more efficient public key based protocols.
\subsection{Protocol}
\begin{figure}
    \centering
    \includegraphics[width=0.7\textwidth]{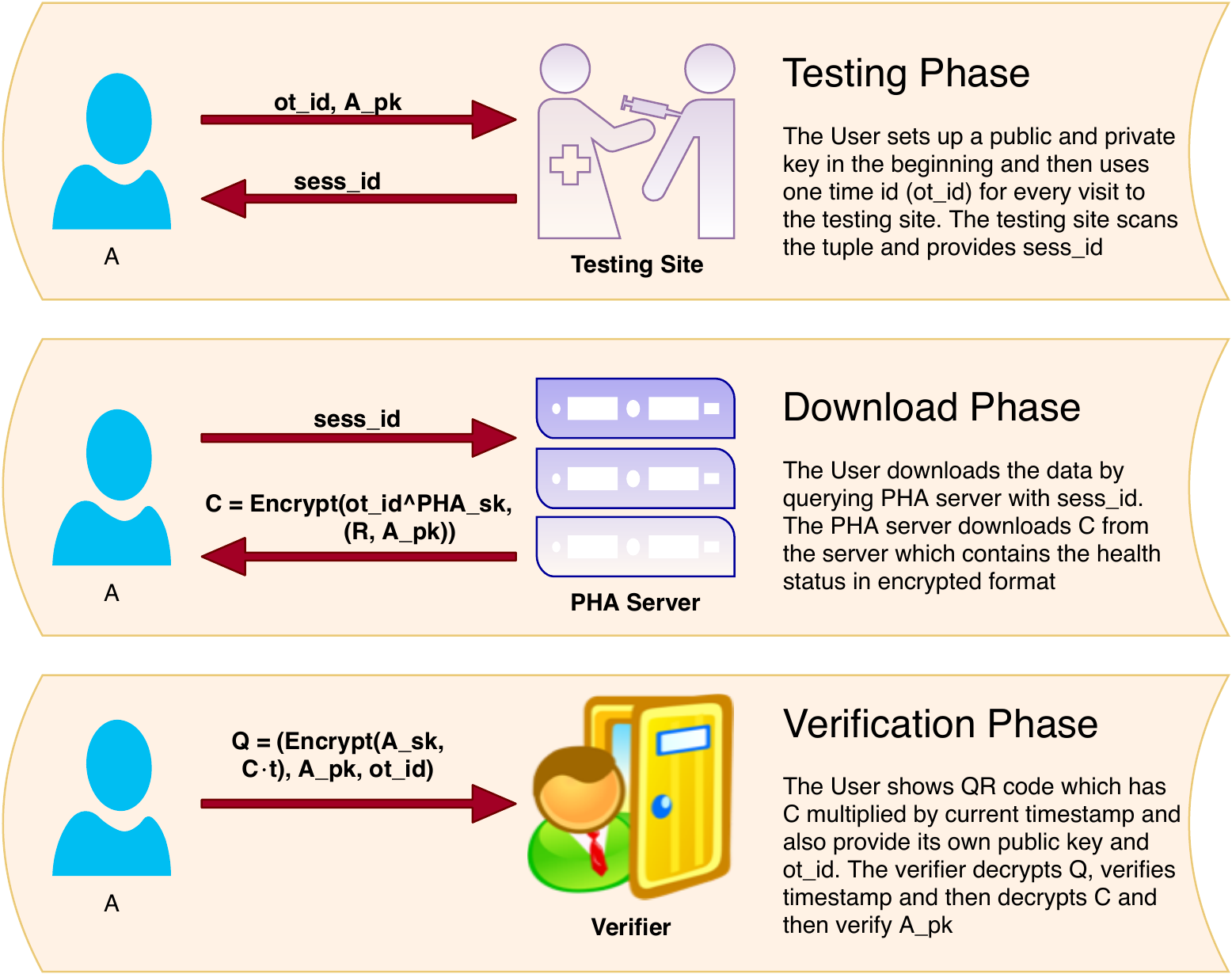}
    \caption{\textbf{Protocol description}: The above user journey diagram illustrates the protocol and message exchanges. Note that the protocol assumes that the \textit{testing site} and the \textit{PHA server} coordinate together to obtain every tested user's information and process it accordingly.}
    \label{fig:protocol}
\end{figure}

\paragraph{Setup phase}
\begin{itemize}
    \item The \textit{issuer} PHA performs $\KeyGen$ operation to obtain $\phapk$ and $\phask$. $\phapk$ can be released publicly or given to business owners in a secure manner based on the access control policy on who can do verification.
    \item The \textit{holder} A performs $\KeyGen$ operation to gather $\Apk$ and $\Ask$. $\Ask$ is securely stored in a trusted enclave which even A can not access through system layer security but can only be used by authorized app to perform encryption.
\end{itemize}

\paragraph{Test phase}
\begin{itemize}
    \item A goes to a testing station and generates a one time identifier $ot\_id$ and provides a QR-code containing information $(ot\_id, \Apk)$
    \item Testing site generates a unique $sess\_id$ for A and provide it to them through QR-code or any other message exchange service. The testing site locally stores the tuple $(ot\_id, \Apk, sess\_id)$.
    \item This $sess\_id$ is used to carry the testing specimen around and used as the primary identifier for the testing result as well as specimen.
\end{itemize}

\paragraph{Upload phase}
\begin{itemize}
    \item Once the testing result data $R$ is ready for the user $A$ the testing site sends $(sess\_id, ot\_id, \Apk, R)$ to the PHA. Based on the result, the PHA can take additional action like contact tracing and etc. or upload the result to the PHA server described as follows.
    \item The PHA uploads the following encrypted result $C=\Encrypt(\phask \oplus ot\_id, (R,A_{pk}))$ and hosts it as a tuple $(sess\_id, C)$.
\end{itemize}

\paragraph{Download phase}
\begin{itemize}
    \item A downloads $C$ by sending $sess\_id$ to the server.
    \item A performs the following computation to decrypt the result $R, A_{pk} = \Decrypt(\phapk \oplus ot\_id, C)$
    \item $A$ verifies $\Apk$ received from the server and interprets the result.
\end{itemize}

\paragraph{Verification phase}

\begin{itemize}
    \item $A$ computes time based signature $S$ by performing $S = encrypt(C\cdot t, \Ask)$ where $t$ is the current timestamp, rounded off to account for clock drift.
    \item $A$ computes QR code as follows $Q=(S, ot\_id, \Apk)$ and shows it to the verifier $V$
    \item $V$ obtains $(S, ot\_id, \Apk) = Q$
    \item $V$ obtains $C\cdot t = \Decrypt(id\oplus \phapk, S)$
    \item $V$ generates $t$ locally and computes its multiplicative inverse $t^{-1}$ to obtain $C$
    \item $V$ verifies the identity of $A$ by verifying the signature $S = \Decrypt(\Apk,C\times t)$
\end{itemize}

%% file: tex/analysis.tex
\section{Analysis}\label{sec:analysis}
\subsection{Security}
The overall security of the protocol not only depends on the security offered by the cryptographic mechanisms but also on the assumption that a user $A$ can not share its secret key $sk$ with other users otherwise other users can use the same key to use $A$'s \textit{health status}. This is a reasonable assumption given that key management can be performed these days on smartphone using secure enclaves~\cite{ios_secure_enclave}. In the Section~\ref{sec:req} we discuss various functional and privacy requirements of the protocol. P1 and P3 are enforced by making sure that $C$ is not known to any curious user unless they hold $\Apk$ as well as $ot\_id$ for the given record. P2 requirement is met by ensuring that the \textit{health status} $R$ is not tied to any other additional health of the user. For practical purposes, the \textit{issuer} might want to communicate more health information to the user which could be performed by appending an additional layer of encryption to sensitive set of health information that does not need to be shared with the verifier. P4 requirement is attained through the secrecy and ephemeral nature of the $ot\_id$ and $sess\_id$.
\subsection{Computation Efficiency}
The computation in this protocol does not involve any heavyweight computation and only relies on a single public key encryption decryption by the $holder$ for every verification round. Since the verification is supposed to happen only when a user visits a public place or business, two subsequent computation will have a sufficient time window between them making it practically possible. For $n$ users the \textit{issuer} has to perform $n$ encryption that requires two exponentiation operation for el-gamal~\cite{elgamal} and decryption requires one exponentiation. Exponentiation operation for $m$ bits of data requires complexity of $\mathcal{O}(m^{log_23})$ under Karatsuba algorithm. Furthermore, some of this exponentiation operations could be pre-computed because of the el-gamal algorithm and all $n$ encryptions are independent of each other making the protocol parallelize at a linear scale.
\subsection{Communication Efficiency}
For $m$ bits of information, the key expansion of el-gamal~\cite{elgamal} algorithm used in the proposed protocol would introduce the message complexity of $2\cdot m$. As shown in the Figure~\ref{fig:protocol}, message exchange mostly involves exchange of identifiers and ciphertext, hence, communication is always linear with the size of message $m$ and small enough to be communicated with a single QR-code. In the message exchange performed in the verification phase, the total bits required for messaging can be given as $2\cdot m + |\Apk| + |ot\_id|$ where $|\Apk|$ refers to the size of the public key which is typically $2048$ bits and $|ot\_id|$ refers to the one time identifier which is usually $256$ bits. Overall, this information can fit within a QR-code and hence can be communicated from \textit{holder} to \textit{verifier} in a single round. In addition to this, our protocol requires minimum interaction with an online server by mandating functional requirements F1 and F2.

%% file: tex/discussion.tex
\section{Discussion}
\subsection{Unintended Consequences} The idea of restricted access based on either a negative test result or . The letter~\cite{noauthor_2020-05-13_2020,schwartz_no_2020} from EFF and ACLU California highlights several criticism for the California's bill~\cite{noauthor_bill_nodate} for the use of verifiable credential technology~\cite{W3C} to develop immunity passport.
\paragraph{Privacy leakage}
Most of the privacy concerns have been discussed in the section~\ref{sec:analysis} and one of the central point for leaking privacy and bypassing the verification security is the secrecy of the secret key of the \textit{holder}. This is possible to obtain if the phone is jailbroken or rooted which might be the case for some of the phones especially the ones with old unupdated operating systems. Furthermore, this could encourage the work in attacking the system layer of security, however, it has been becoming increasingly difficult with the latest operating systems and the app can enforce a minimum version of operating system to enforce higher security.\\
The other possibility with this protocol is that the result $R$ of a user is shared permanently with the verifier which might not be a good idea if the result $R$ contains other sensitive information about the user as well. Hence, in our proposed protocol, we keep it limited to only the test result of a user that does not hold much of significance once the pandemic is over.
\paragraph{Equity}
The proposed protocol relies on the smartphones that inevitably would lead to equity concerns. This would be even more substantiated in the developing countries. Hence we want the stakeholders to be cognizant of this aspect of the proposed protocol. The ongoing direction of this work is to adjust the verification scheme such that it is supported by a paper based QR-code which can be given to the users who do not have access to the smartphones.
\paragraph{Behavioral changes}
One of the major unintended consequence of a \textit{health status based access} could be to attain immunity through risky pathways. Such a behavioral change could lead to more infection spread and fatalities due to more number of people exposing themselves in order to get access to public places for their own self-interest or financial reasons. This is one of the unintended consequence of the restricted access control that can not be addressed entirely by the proposed technology.
\subsection{Effectiveness of restricted access}
Notwithstanding the above mentioned points, this work does not advocate for restricting access to individuals who are not able to present their testing results due to different reasons. Rather, such a system could be used for data collection and retrospective analysis for future interventions. One such example is public places increasing their risk level if the total percentage of people entering the venue and showing the \textit{digital passport} decline. A majority of the community discussion around the restricted access has been on a binary level of whether to use it completely or not at all. We hope that such systems can be introduced gradually and mindfully to the existing ecosystem such that it allows decision making in data-driven manner and allow the technology to mature.

%% file: tex/conclusion.tex
\section{Conclusion}\label{sec:conclusion}
In this work, we present a protocol for testing verification for curbing the COVID-19 disease spread by restricting the access to only recently tested individuals. The focus of this work is not on the actual deployment but rather protocol design that circumvents some of the practical issues which existing systems have. We also discuss various unintended consequences of such a platform and in what ways it can be addressed. This work does not advocate for immunity passports but rather verification of testing results that would encourage testing and reduce the chance of infectious individuals entering a public place.